\documentclass{article}

\usepackage{epsfig}
\usepackage{psfrag}
\usepackage{chicago}
\usepackage{amsmath}

\renewcommand{\vec}[1]{\boldsymbol{\rm #1}}

\newcommand{\rhosand}{\rho_{\text{sand}}}
\newcommand{\rhoair}{\rho_{\text{air}}}
\newcommand{\rhoquartz}{\rho_{\text{quartz}}}

\newcommand{\veff}{v_{\text{eff}}}

\begin{document}

\title{Modelling transverse dunes}
\author{Veit Schw\"ammle$^{(1)}$ and Hans J. Herrmann$^{(1)}$}


\maketitle

\begin{abstract}
Transverse dunes appear in regions of mainly unidirectional wind and high sand availability. 
A dune model is extended to two dimensional calculation of the shear stress. It is applied to 
simulate dynamics and morphology of transverse dunes which seem to reach translational invariance and 
do not stop growing. Hence, simulations of two dimensional dune fields have been performed. 
Characteristic laws were found for the time evolution of transverse dunes. 
Bagnold's law of the dune velocity is modified and reproduced. The interaction between transverse dunes
led to interesting results which conclude that small dunes can pass through bigger ones.
\end{abstract}

\section{Introduction}

Transverse dunes have been investigated with constant interest recently. Field measurements 
have been carried out in order to obtain more knowledge about airflow and sediment transport 
over transverse dunes \cite{Wilson72,Lancaster82,Lancaster83,Mulligan88,Burkinshaw93,McKenna2000}. 
Several numerical models have been proposed to model dune formation \cite{Wippermann86,zeman-jensen:88,Fisher88,Stam97,NishimoriXX,boxel-arens-van_dijk:99,van_dijk-arens-boxel:99,herrmann-sauermann:2000,MomijiWarren2000,SauermannKroy2001,KroySauermann2002}. But still many questions are not answered 
and some efforts are needed to better understand dune morphology and dynamics or even to
anticipate dune formation. The simulations in this article will try to answer some 
questions and justify the importance of further field measurements.

First some general aspects
will be introduced. A model for dunes in three dimensions will show why translational invariance 
appears in transverse dune formation. The time evolution and the dune velocity in 
a two dimensional model with constant sand influx will be presented.  
The time evolution of a model of two dimensional dunes with periodic boundary conditions will give some 
more insight into dune field dynamics. 
A final discussion of the statement that dunes behave like
solitons will close this article. 

\section{General aspects of transverse dunes}
\label{sec:trans_intro}

About 40\% of all terrestrial sand seas are covered by transverse dunes. They are mostly
located in sand seas where sand is available. Thus in larger sand seas one 
mainly finds ensembles of many transverse dunes which interact with respect 
to their dynamics. The crest to crest
spacing ranges from a few meters to over 3 km \cite{Breed79}.
They are common in the Northern Hemisphere, for example in China, and along coasts.
On Mars transverse dunes dominate the sand seas. 
\begin{figure}[h]
  \begin{center}
    \includegraphics[width=0.8\textwidth,angle=90]{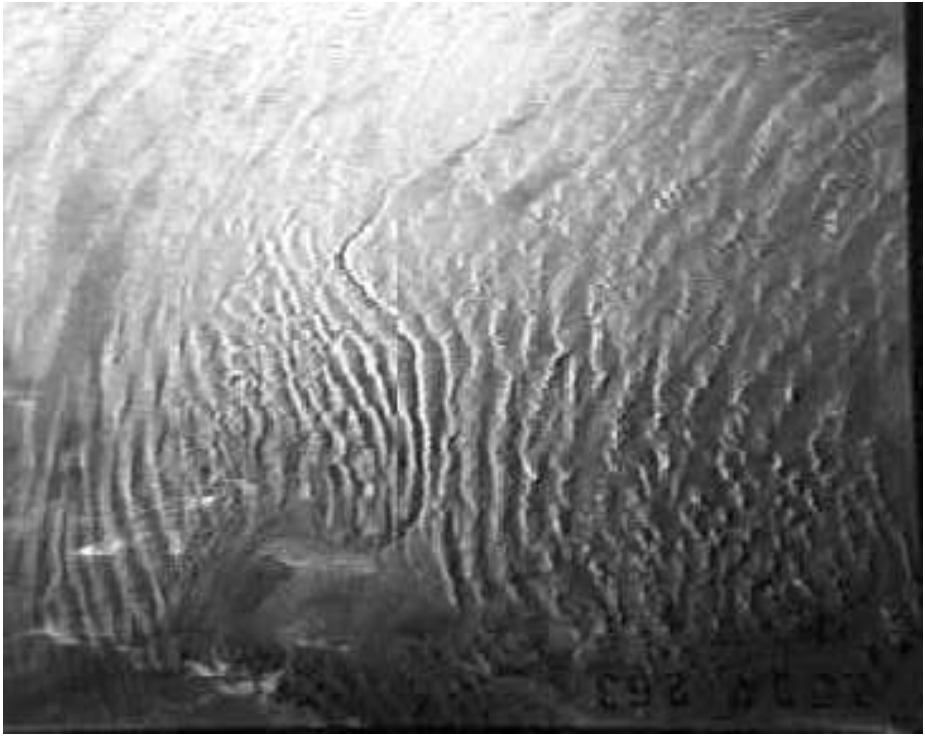}
    \caption{An aerial photo (STS047--153A--263) of transverse dunes in Lop Nur, China.
Image courtesy
        of Earth Sciences and Image Analysis Laboratory, NASA
        Johnson Space Center (http://eol.jsc.nasa.gov)}
    \label{fig:transv_photo}
  \end{center}
\end{figure}
An example of a field of transverse dunes in China is shown in Figure~\ref{fig:transv_photo}. 
Normally transverse dunes have more irregular patterns and even smaller
hierarchies of smaller transverse dunes can be found in a field of big transverse dunes.
Strong winds coming essentially from the same direction are the main environment where 
this type of dune can be found. \citeN{Cooke93} 
proposed that in highly mobile environments cross winds distort the dune shape.
\begin{figure}[h]
  \begin{center}
    \includegraphics[width=0.8\textwidth]{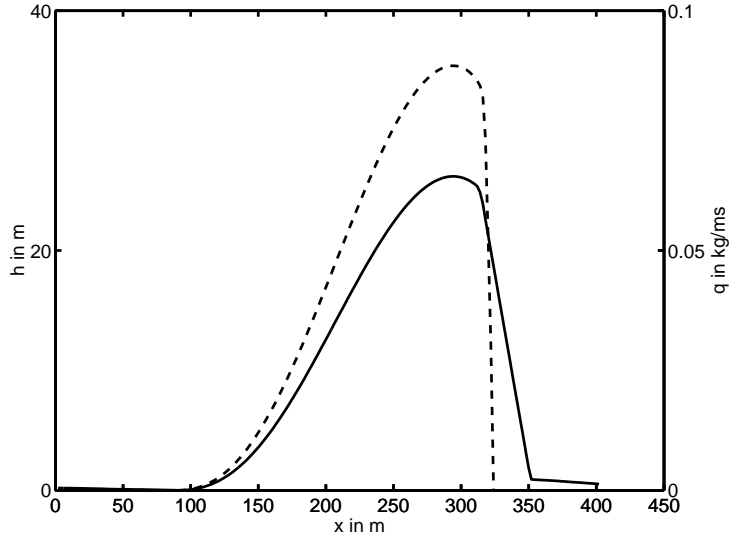}
    \caption{Two dimensional cut through a single transverse dune profile $h(x)$ (solid line) and the sand 
             flux $q(x)$ (dashed line) over it. The 
             wind is coming from the left.}
    \label{fig:transv_single}
  \end{center}
\end{figure}

In Figure~\ref{fig:transv_single} a dune of a height of about $27$ m is depicted which is 
part of a  two dimensional calculation of our model. This dune is situated between other
similar dunes which together result in a simulation of a dune field of a length of four 
kilometers. The dashed curve shows the sand flux which is set to zero at the slip face where
the shear stress is not strong enough to entrain sand. This occurs in the region of flow separation, 
defining the boundary between quasi-laminar flow and the turbulent layer of the eddy after
the brink. 
The brink separates windward side and slip face. In this case the brink and the crest do not 
coincide which means in this case that this dune has not yet reached a stationary state. 

\section{The dune model}
\label{sec:model}

The model described here can be seen as a minimal model including the main processes
of dune morphology. 
As predecessor of the model described here the work of \citeN{SauermannPhD2001} revealed 
interesting new insights into dynamics and 
formation. The model is now extended to a  two dimensional shear stress calculation 
(longitudinal and lateral direction), a full sand bed and different boundary conditions.
Nevertheless the wind is restricted to be constant and unidirectional in time. 
In this article the model is used to simulate transverse dune fields.
In every iteration the horizontal shear stress $\vec{\tau}$ of the wind, the saltation flux $\vec q$ 
and the flux due to avalanches are calculated. The time scale of these processes is much smaller than the 
time scale of changes in the dune surface so that they are treated to be instantaneous. In the 
calculation of the surface evolution a time step of 3--5 hours is used. In the following the
different steps at every iteration are explained. 

\paragraph{The air shear stress $\tau$ at the ground:}

The shear stress perturbation over a single dune or over a dune field is calculated using
the algorithm of \citeN{weng-etal:91}. The $\tau_x$-component points in wind direction and the 
$\tau_y$-component denotes the lateral direction. The calculation is made in Fourier space, where 
$k_x$ and $k_y$ denote the wave numbers,
\begin{eqnarray}
  \label{eq:tau_x}
  \hat\tau_x(k_x,k_y) = 
    \frac{h(k_x,k_y) k_x^2}{|k|} 
  \frac{2}{U^2(l)} \cdot  \nonumber \\
  \left( 1 +   
   \frac{2 \ln L|k_x| + 4 \gamma + 1 +
      i \, \text{sign}(k_x) \pi}{\ln l/z_0} \right),
\end{eqnarray}
and
\begin{equation}
  \label{eq:tau_y}
  \hat\tau_y(k_x,k_y) = \frac{h(k_x,k_y) k_x k_y}{|k|}
  \frac{2}{U^2(l)},
\end{equation}
where  $|k| = \sqrt{k_x^2 + k_y^2}$ and $\gamma=0.577216$ 
(Euler's constant). $U(l)$ is the normalized velocity of the undisturbed logarithmic profile at 
the height of the inner region $l$ \cite{SauermannPhD2001} defined in \citeN{weng-etal:91}. 
The roughness length $z_0$ is set to 
$0.0025$ m and the so called characteristic length $L$ to $10$ m.

\begin{figure}[h]
  \begin{center}
    \includegraphics[width=0.8\textwidth]{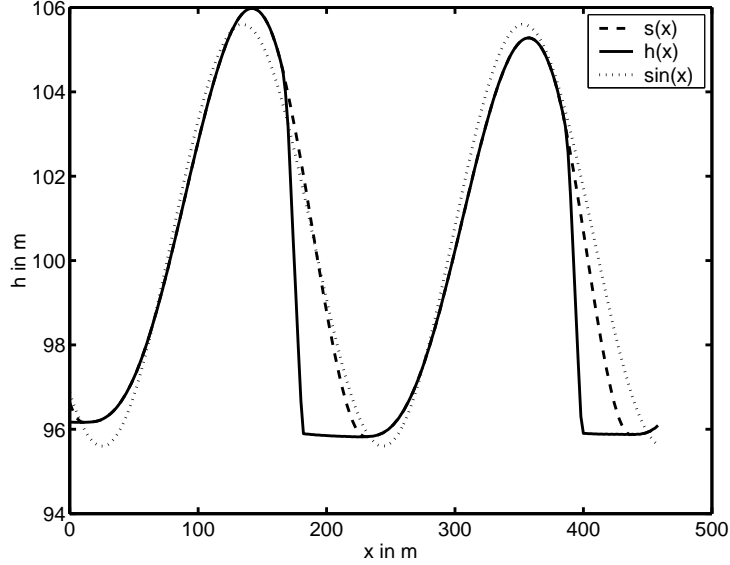}
    \caption{The transverse dune profile $h(x)$, its separation bubble $s(x)$ and a sine 
             function. The separation bubble ensures a smooth surface. 
             The wind is coming from the left.}
    \label{fig:transv_sinus}
  \end{center}
\end{figure}

Equations~(\ref{eq:tau_x}) and (\ref{eq:tau_y}) are calculated in Fourier space and have to be 
multiplied with the logarithmic 
velocity profile in real space in order to obtain the total shear stress. The surface is 
assumed to be instantaneously rigid and the effect of 
sediment transport is incorporated in the roughness length $z_0$. 
For slices in wind direction the separation streamlines in the lee zone
of the dunes are fitted by a polynomial of third order going through the point of the brink 
$x_0$. The length of the separation streamlines is determined by allowing a maximum slope of $14^0$ 
\cite{SauermannPhD2001}. Where the 
separation streamlines cross the surface at the reatachment point $x_1$, a new stream line is 
calculated as third polynomial attaching $x_0$ and $x_1$. The separation bubble guarantees a smooth 
surface and the shear stress in the area of the separation bubble is set equal to zero. 
Problems can arrive due to numerical fluctuations in the value of the slope of the brink where 
the separation bubble begins and its influence on the calculation of a separation streamline for each 
slice. To get rid of this numerical error the surface is Fourier-filtered by cutting the small frequencies. 
Figure~\ref{fig:transv_sinus} depicts the separation bubble and an interesting
similarity of the separation bubble profile to a sine function. 

\paragraph{The saltation flux $q$:}

The time to reach the steady state of sand flux over a new surface
is several orders of magnitude smaller than the time scale of the surface evolution.
Hence, the steady state is assumed to be reached instantaneously. The length scale of the
model is too large to include sand ripples. Nevertheless the kinetics and the characteristic
length scale of saltation influence the calculation by breaking the
scale invariance of dunes and by determining the minimal size of a barchan dune \cite{SauermannKroy2001}. 
A calculation of the saltation transport by the well known flux relations 
\cite{Bagnold41,Lettau78,Sorensen91} would restrict the model to saturated sand flux which
is not the case for example at the foot of the windward side of a barchan dune due to little
sand supply or at the end of the separation bubble in the interdune region between 
transverse dunes due to the vanishing shear stress in the separation bubble.
The sand density $\rho(x,y)$ and the grain velocity $\vec u(x,y)$ are integrated in vertical direction and
 calculated from mass and momentum conservation, respectively. We simplified the closed model of 
\cite{SauermannKroy2001} by neglecting the time dependent terms and the convective term of the grain 
velocity $\vec u(x,y)$. The expansion to
two dimensions yields Equations~(\ref{eq:3d_rho}) and (\ref{eq:3d_u}) where $\rho$ and $\vec u$ are
determined from the before obtained shear stress and the gradient of the actual surface, 
\begin{equation}
  \label{eq:3d_rho}
  \text{div} \, (\rho \, \vec u) 
  = \frac{1}{T_s} \rho \left( 1 - \frac{\rho}{\rho_s} \right) \; 
  \begin{cases}
    \Theta(h) & \rho < \rho_s\\
    1         & \rho \ge \rho_s
  \end{cases}
  ,
\end{equation}
with
\begin{equation}
  \label{eq:rho_s_tau}
  \rho_s = \frac{2 \alpha}{g} \left( |\vec \tau| - \tau_t \right) \quad \quad
  T_s = \frac{2 \alpha | \vec u|}{g} \, \frac{\tau_t}{\gamma (|\vec \tau| - \tau_t)}.
\end{equation}
and
\begin{equation}
  \label{eq:3d_u}
   \frac{3}{4} \, C_d \frac{\rho_{\text{air}}}{\rho_{\text{quartz}}} d^{-1} \, (\vec \veff - \vec 
u)|\vec \veff - \vec u|
    - \frac{g}{2 \alpha} \frac{\vec u}{|\vec u|}
    - g \, \vec \nabla \, h
    = 0,
\end{equation}
where $v_{\text{eff}}$ is the velocity of the grains in the saturated state,\\
\begin{equation}
 \vec v_{\text{eff}} =   \frac{2 \vec u_*}{\kappa |\vec u_*|} \cdot \nonumber 
\end{equation}
\begin{equation}
  \label{eq:3d:v_eff_of_tau_g0}
\left( \sqrt{\frac{z_1}{z_m} u_*^2 + 
       \left( 1 - \frac{z_1}{z_m} \right)  
      \, u_{*t}^2} 
     + \left( 
     \ln{\frac{z_1}{z_0}} -2 \right) \, \frac{u_{*t}}{\kappa} 
  \right),
\end{equation}
and
\begin{equation}
  \label{eq:ustar_s}
  u_* = \sqrt{\tau / \rhoair}
\end{equation}
The constants and model parameters have been taken from \cite{SauermannKroy2001} and are 
summarized here:  $g=9.81\,$m$\,$s$^{-2}$,$\kappa=0.4$,
$\rhoair=1.225\,$kg$\,$m$^{-3}$,
$\rhoquartz=2650\,$kg$\,$m$^{-3}$, $z_m=0.04\,$m,
$z_0=2.5~10^{-5}\,$m, $D=d=250\,\mu$m, $C_d=3$,
$u_{*t}=0.28\,$m$\,$s$^{-1}$, $\gamma=0.4$,$\alpha=0.35$ and $z_1=0.005\,$m.
The sand density and the sand velocity define the sand flux over a surface element
$\vec q(x,y)=\vec u(x,y) \rho(x,y)$. 

\paragraph{Avalanches:}

Surfaces with slopes which exceed the maximal stable angle of a sand surface, the called 
{\em angle of repose} $\Theta \approx 34^o$, produce avalanches which slide down in the 
direction of the steepest descent. The unstable surface relaxes to a somewhat smaller
angle. For the study of dune formation two global properties are of interest. 
These are the sand transport downhill due to gravity and the maintenance of the angle of
repose. To determine the new surface after the relaxation by avalanches  the model
proposed by \citeN{Bouchaud94} is used, 
%
\begin{equation}
  \label{eq:dhdt_a}
  \frac{\partial h}{\partial t} = -
  C_a R \left( \left| \nabla h \right| - \tan \Theta \right)
\end{equation}
\begin{equation}
  \label{eq:drho_adt}
  \frac{\partial R}{\partial t} 
  + \nabla \left(R \vec u_a \right) =
  C_a R \left( \left| \nabla h \right| - \tan \Theta \right),
\end{equation}
where $h$ denotes the height of the sand bed, $R$ the height of the moving layer, $C_a$
is a model parameter and the velocity $\vec u_a$ of the sand grains in the moving layer is obtained by,
\begin{equation}
  \label{eq:3d:ua}
  \vec u_a = - u_a \frac{\nabla h}{\tan \Theta},
\end{equation}
where $u_a$ is the velocity at the angle of repose. 
Like in the calculation of the sand flux the steady state of the avalanche model
is assumed to be reached instantaneously. In the dune model a certain amount of sand 
is transported over the brink to the slip face and in every iteration the sand grains
are relaxed over the slip face by this avalanche model determining the steady state. 
\paragraph{The time evolution of the surface}

The calculation of the sand flux over a not stationary dune surface leads to changes 
by erosion and deposition of sand grains. The change of the surface profile can be expressed
using the conservation of mass,
\begin{equation}
   \partial_t \rho + \nabla \Phi = 0 \, ,
   \label{eq:time_evol_h}
\end{equation}
where $\rho$ is the sand density and $\Phi$ the sand flux per time unit and area. Both
$\rho$ and $\Phi$ are now integrated over the vertical coordinate assuming that the 
dune has a constant density of $\rho_{\text{sand}}$,
\begin{equation}
  \label{eq:height_rho}
  h = \frac{1}{\rhosand} \int \rho dz , \quad \quad
  \vec q = \int \vec \Phi dz.
\end{equation}
Thus Equation~(\ref{eq:time_evol_h}) can be rewritten, as
\begin{equation}
  \label{eq:masscons_h}
  \frac{\partial h}{\partial t} = - \frac{1}{\rhosand} \nabla{\vec q}.
\end{equation}
Finally, it is noted that Equation~(\ref{eq:masscons_h}) is the only remaining time dependent
equation and thus defines the characteristic time scale of the model which is normally
between 3--5 hours for every iteration.

\paragraph{The initial surface and boundary conditions:}

As initial surface we take a plain sand bed of arbitrary sand height over the solid ground.
The surface can additionally be disturbed by small Gaussian hills.
An initial surface has to be smooth (at least under consideration of the separation bubble) 
and have slopes not larger than the angle of repose.
The boundary conditions influence the surface 
height $h$ with its separation bubble, the sand flux $q$ and the height $R$ of the moving 
layer of the avalanche model. The boundary in both directions $x$ and $y$ with respect to the 
direction of the incoming wind are open or periodic. 
At open boundary in $x$--direction an additional parameter controls the sand influx $q_{in}$ into
the simulated dune field. It is set constant along the lateral direction at $x=0$.

All calculations presented in the following sections 
are made with the conditions of a completely filled sand bed and unidirectional 
wind. All simulations model dune fields instead of single dunes.

\section{The model of three dimensional dunes and translational invariance}
\label{sec:trans_2dim}

The main aim in this section is to justify why models of two dimensional dunes can be used 
in the  following sections. The advantage to omit the lateral dimension makes it possible to
look at larger dune fields within a still tolerable cost of computional time. 

A plain initial surface would lead to no change of the height profile. This is because
the system needs at least one small fluctuation to begin dune growth. Therefore as initial
surface a large number of low Gaussian hills is introduced.
\begin{figure}[htb]
  \begin{center}
    \includegraphics[width=0.8\textwidth]{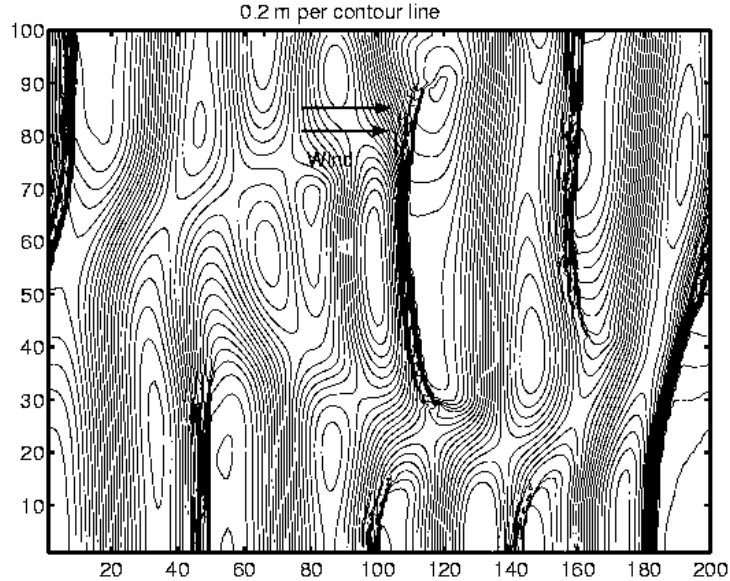}
    \caption{Surface after $1.49$ years. The shear velocity is $u_*=0.45 $ m s$^{-1}$, the 
             boundary conditions are periodic in wind direction and open in lateral 
             direction. Some slip faces can be seen. One unit corresponds to the 
             length of $2$ m.}
    \label{fig:transv_2d2}
  \end{center}
\end{figure}
The simulation models dune dynamics for a dune field of a length and width of $400$ m and 
$200$ m, respectively.
The boundary conditions are periodic in wind direction and open in lateral direction and 
the shear velocity is $u_*=0.45 $ m s$^{-1}$. First the Gaussian hills lead to a 
growth at their corresponding positions on the dune field. After some time they build a slip face
which extends its size in lateral direction. 

\begin{figure}[tbp]
  \begin{center}
    \includegraphics[width=0.8\textwidth]{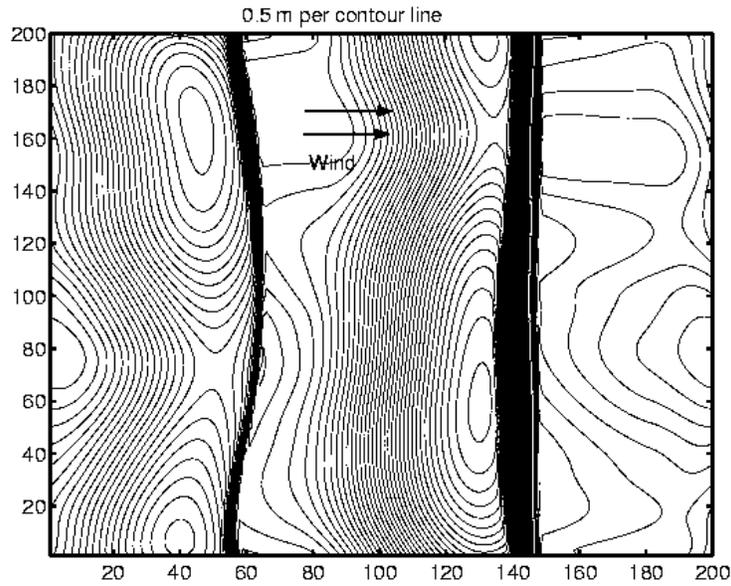}
    \caption{Surface of a transverse dune field with periodic boundary conditions in both 
             directions $9.5$ years after initiation. The system reached a state close
             to translational invariance. The shear velocity is $u_*=0.4$ m s$^{-1}$. One unit 
             corresponds to the length of $2$ m.}
    \label{fig:transv_2d6}
  \end{center}
\end{figure}

We assume that transverse dunes try to reach
a state of translational invariance. For an illustration of this dynamics see 
Figure~\ref{fig:transv_2d2}. The slip faces become wider until they reach the lateral boundary and 
extend over the entire width of the simulated fields. Thus the slip face traps
all the sand going over the brink. The trapped sand relaxes there through avalanching and maintains 
the angle of repose constant. Shear stresses of the wind field transport more sand over the brink
than that which is transported down at the slip face by avalanches.
Hence, transverse dunes are growing wherever there is a part of their lee zone in which
saltation transport can be neglected (Section~\ref{sec:trans_const}). 
When dunes grow the length of their separation bubble increases.
But for a dune field with a limited length due to the periodic boundary condition and 
to a certain number of dunes which increase their mutual distance there is a state where
the number of dunes must decrease by one.
This leads to a displacement of the dune with the lowest height which looses its
sand to the next dune situated upwind. In this state
the system breaks the symmetry of translational invariance and a part of the slip face disappears.
There sand is transported to the following dune by saltation. When the dune has vanished
once again the system approaches translational invariance. Hence, the effect of converging
dunes perturbs the steady growth of transverse dunes in a field with a periodic boundary.
\begin{figure}[bp]
  \begin{center}
    \includegraphics[width=0.8\textwidth]{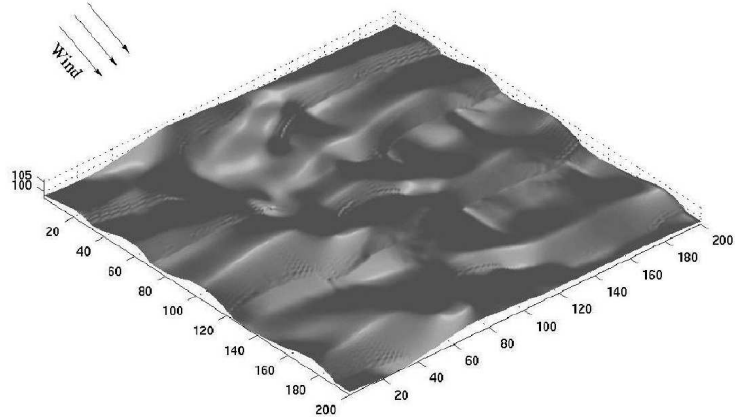}
    \caption{Surface of a transverse dune field with periodic boundary conditions in both 
             directions $1.19$ years after initiation. The shear velocity is 
             $u_*=0.4 $ m s$^{-1}$. Height units are in meters, length and width units in two meters.}
    \label{fig:transv_2d4}
  \end{center}
\end{figure}
To get more information how a system of transverse dunes shows translational invariance
a simulation of a dune fiel with periodic boundary in both horizontal directions 
is made. The field extends over $400$ m in length and width. As initial surface also 
small Gaussian hills are used. Figure~\ref{fig:transv_2d4} shows the height profile after
5,000 iterations which corresponds to a time of $1.19$ years. The results of this simulation
yield the same conclusions as the simulation with open lateral boundary. Also each
merging of two dunes leads to a breaking of the symmetry. Figure~\ref{fig:transv_2d6} shows 
the system at a state close to translational invariance. The final state
of the calculation is reached when only one dune is left. A simulation with 
periodic boundaries in both directions shows no state where the system has a structure
or oscillations in lateral direction.
The conclusion from these calculations for three dimensional dunes would be that an open system 
of transverse sand dunes reaches translational invariance under ideal 
unidirectional wind. In the precedent simulations the periodic boundary inhibited 
the system to break the invariance. A calculation with open boundary conditions in both
horizontal directions which is closer to real dune fields would give more information about 
this assumption but is rather complicated due to the fact that the dunes move out of the 
simulated area. More insight is given in the following section. 
Assuming translational invariance simulations of fields of  two dimensional dunes are much more 
effective. They consume much less computional time and give the opportunity to simulate 
larger dune fields. In the following two sections we consider two situations of two dimensional dunes,
a model with  constant sand influx and a model with periodic boundary. Both 
models lead to new interesting conclusions.

\section{The model of two dimensional dunes with constant sand influx}
\label{sec:trans_const}

The free parameters for this simulation are the sand influx $q_{in}$ and the shear velocity 
$u_*$. As initial surface we choose a plain ground filled with sand because the sand influx
differs at least a little bit from the saturation flux on the dune field. Thus dune formation
is initiated at the beginning of the dune field, i.e. where wind comes in. 

\subsection{Time evolution}
\label{subsec:time_evol_2dc}

\begin{figure}[htb]
  \begin{center}
    \includegraphics[width=0.8\textwidth]{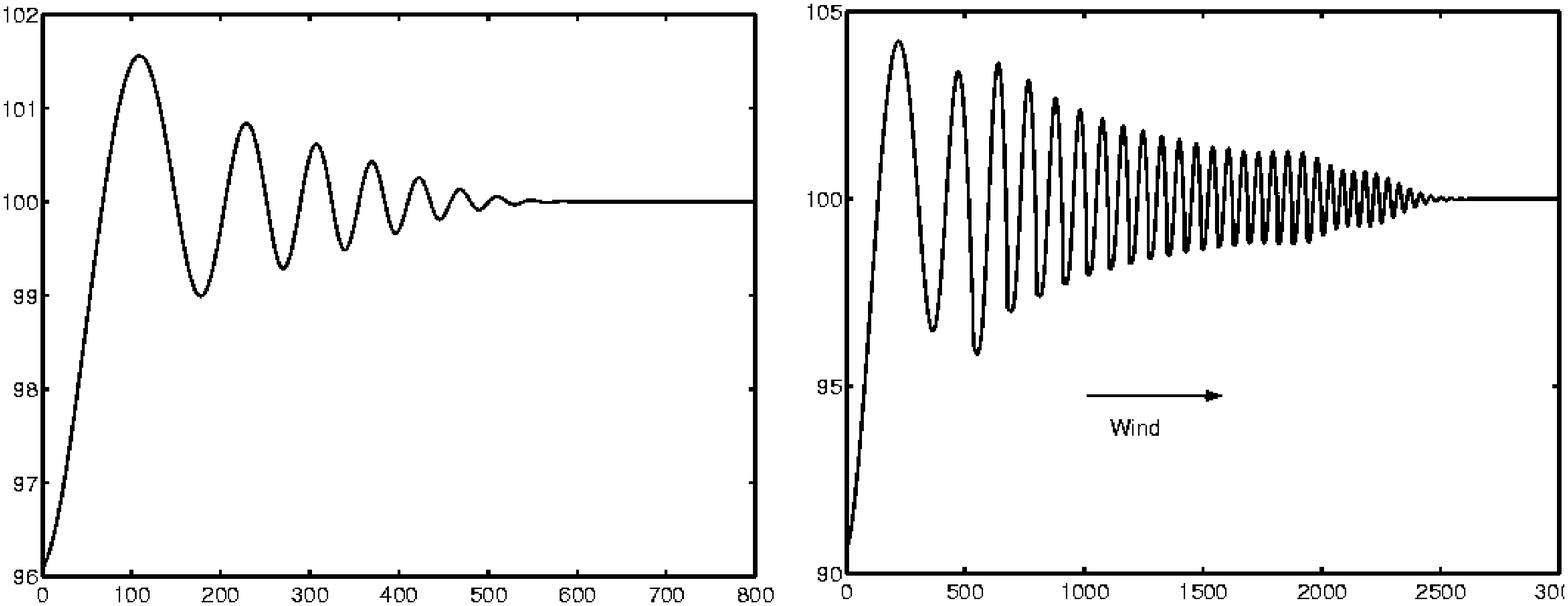}
    \caption{Surface of a two dimensional simulation with constant sand influx after $0.23$ 
             years in the left figure and $1.14$ years on the right. The shear velocity 
             is $u_*=0.5$ m s$^{-1}$ and sand influx is $q_{in}=0.017$ kg m$^{-1}$s$^{-1}$. In 
             the left 
             figure we see that initiation of dune formation began at $x=0$.  Dune height
             decreases with the distance to the inlet of the dune field.}
    \label{fig:transv_evo1}
  \end{center}
%
%
  \begin{center}
    \includegraphics[width=0.8\textwidth]{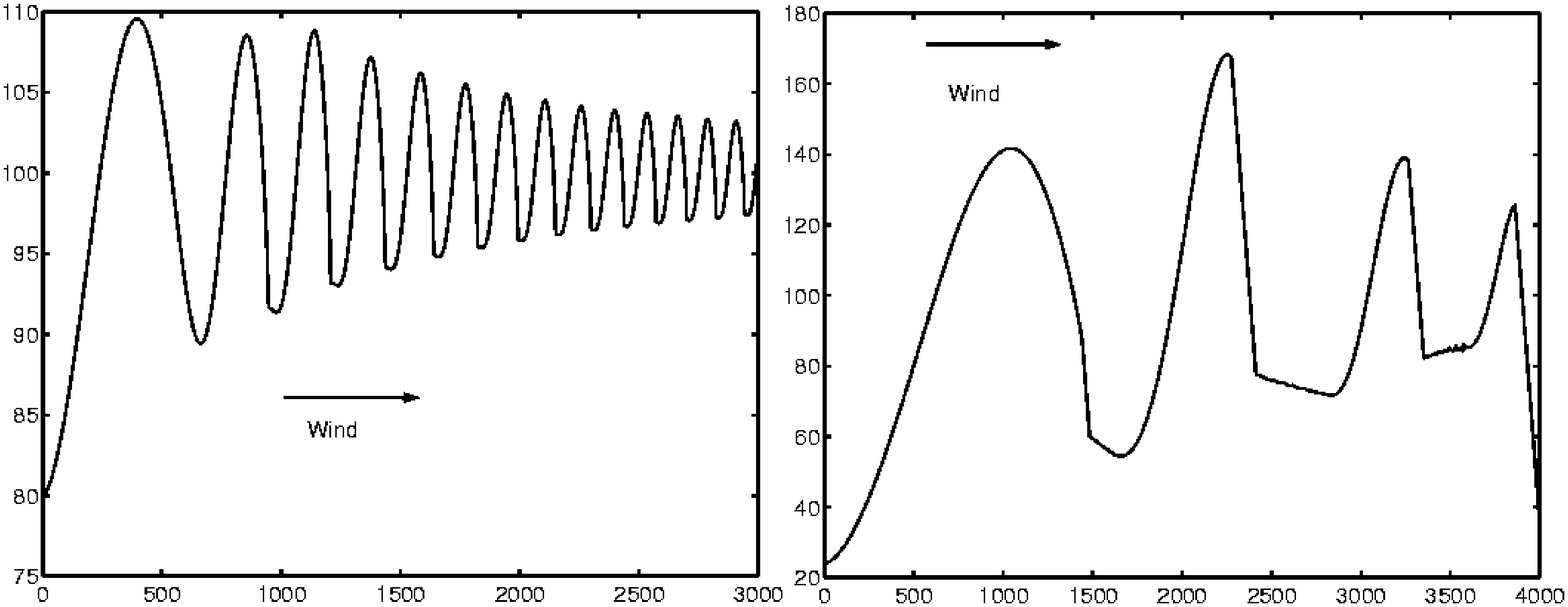}
    \caption{Surface of a two dimensional simulation with constant sand influx after $4.56$
             years in the left figure and $45.58$ 
             years on the right. The shear velocity is $u_*=0.5$ m s$^{-1}$ and sand influx 
             is $q_{in}=0.017$ kg m$^{-1}$s$^{-1}$. On the left the first dune does not have a 
             slip face. A final stationary surface is not reached.}
    \label{fig:transv_evo3}
  \end{center}
\end{figure}

The height profile of a dune field with a length of 4 km is presented at different
times. The sand influx is set $q_{in}=0.017$ kg m$^{1}$s$^{-1}$ and the shear velocity 
$u_*=0.5$ m s$^{-1}$.
A sand influx $q_{in}$ which is not equal to the sand flux of saltation transport over 
a plain surface lowers or raises the inlet of the dune field constantly. This 
initiates a small oscillatory structure which begins to move in wind direction 
(Figure~\ref{fig:transv_evo1}) and generates more and more small dunes. 
The initiating dunes at the inlet of the dune field have increasing size in length and
height. The increasing difference between the 
starting points and the first crest leads to the creation of bigger dunes. With 
increasing size the dunes have a lower velocity $v_{dune}$ (Section~\ref{subsec:dunevel}). 
Hence, 
dune spacing, the distance between adjacent crests, increases with time and 
no dune collides or converges with another. So
real dune fields can maintain a structure of translational invariance without the effect
of the breaking of symmetry which was found in Section~\ref{sec:trans_2dim}. Dune fields 
where the 
sand influx stays rather constant in time can have a more regular structure than dune
fields where the sand influx varies strongly with respect to time.

The slip face of the first dune is missing or is short due to the short evolution time
(Figure~\ref{fig:transv_evo3}). An observation
of this absence for example at a coast where the sea provides a certain amount of sand supply
was not found in the literature. 
The simulation depicted here and any other simulation of dune fields with lengths
of 1 km to 4 km do not show a system that reaches a stationary state.
According to \citeN{Cooke93} older dune fields and less climate changes 
produce larger transverse dunes. The smaller slope at the brink found in this 
modeling agrees also with qualitative observations. In the model a dune height of $100$ m is reached
in roughly $50$ years. Estimates have predicted some $10,000$ years to develop a $100$ m 
dune. Probably this large difference can be explained by a smaller average wind velocity, changes
of wind direction over longer periods and changes in sand supply and climate acting on real
dunes. The so called memory (the time to build a dune beginning with a plain sand bed) 
is related to the ratio of height of the dune and annual rate of sand flux $H/Q_{ann}$
\cite{Cooke93}. The memory can vary by about four orders of magnitude in time.

\begin{figure}[htb]
  \begin{center}
    \includegraphics[width=0.8\textwidth]{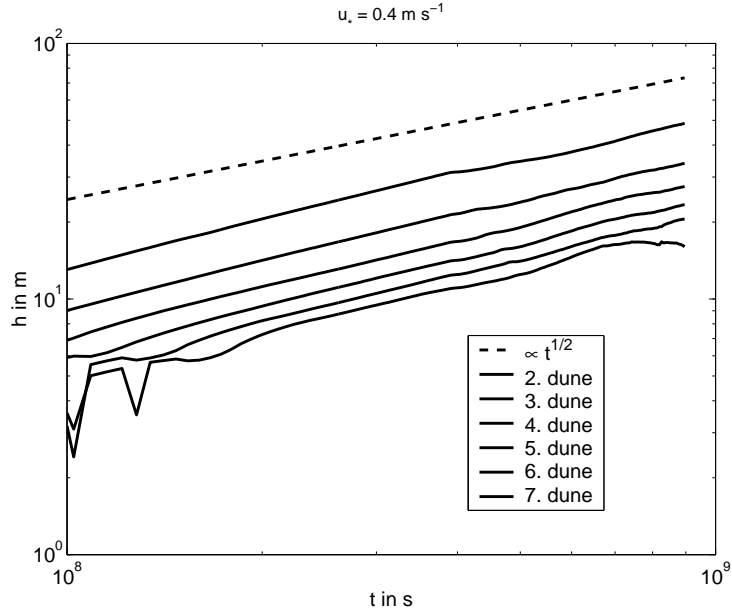}
    \caption{Evolution of the height of the dunes beginning at the position where the wind comes in. 
             The shear velocity is $u_* = 0.4$ ms$^{-1}$. The height increases with the square root of time.}
    \label{fig:transv_scalhq}
  \end{center}
\end{figure}

\begin{figure}[tb]
  \begin{center}
    \includegraphics[width=0.8\textwidth]{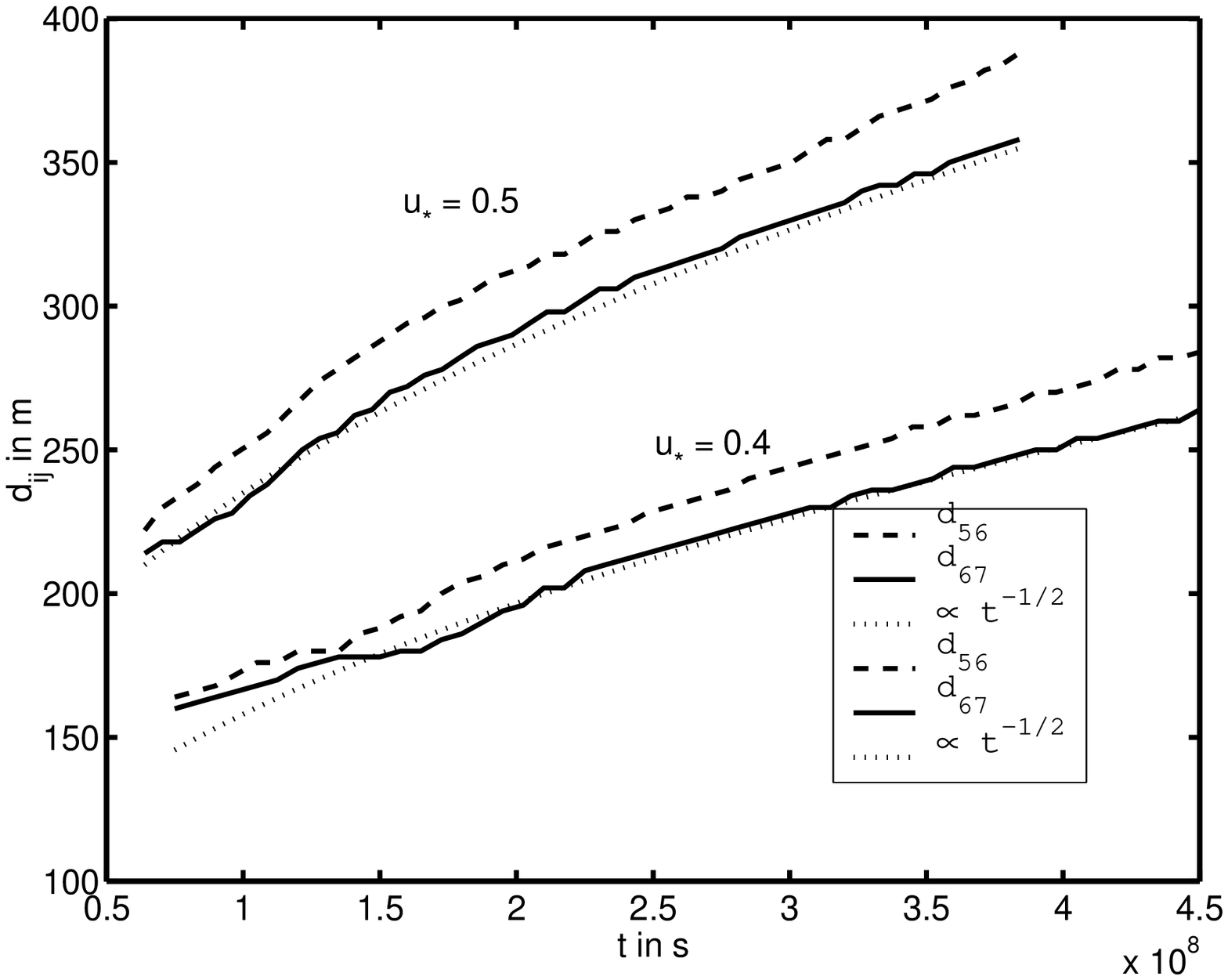}
    \caption{Comparing the evolution of the spacing between dunes for different shear velocities.
             The spacing increases with the square root of time.}
    \label{fig:transv_scalsc}
  \end{center}
\end{figure}

\begin{figure}[tb]
  \begin{center}
    \includegraphics[width=0.8\textwidth]{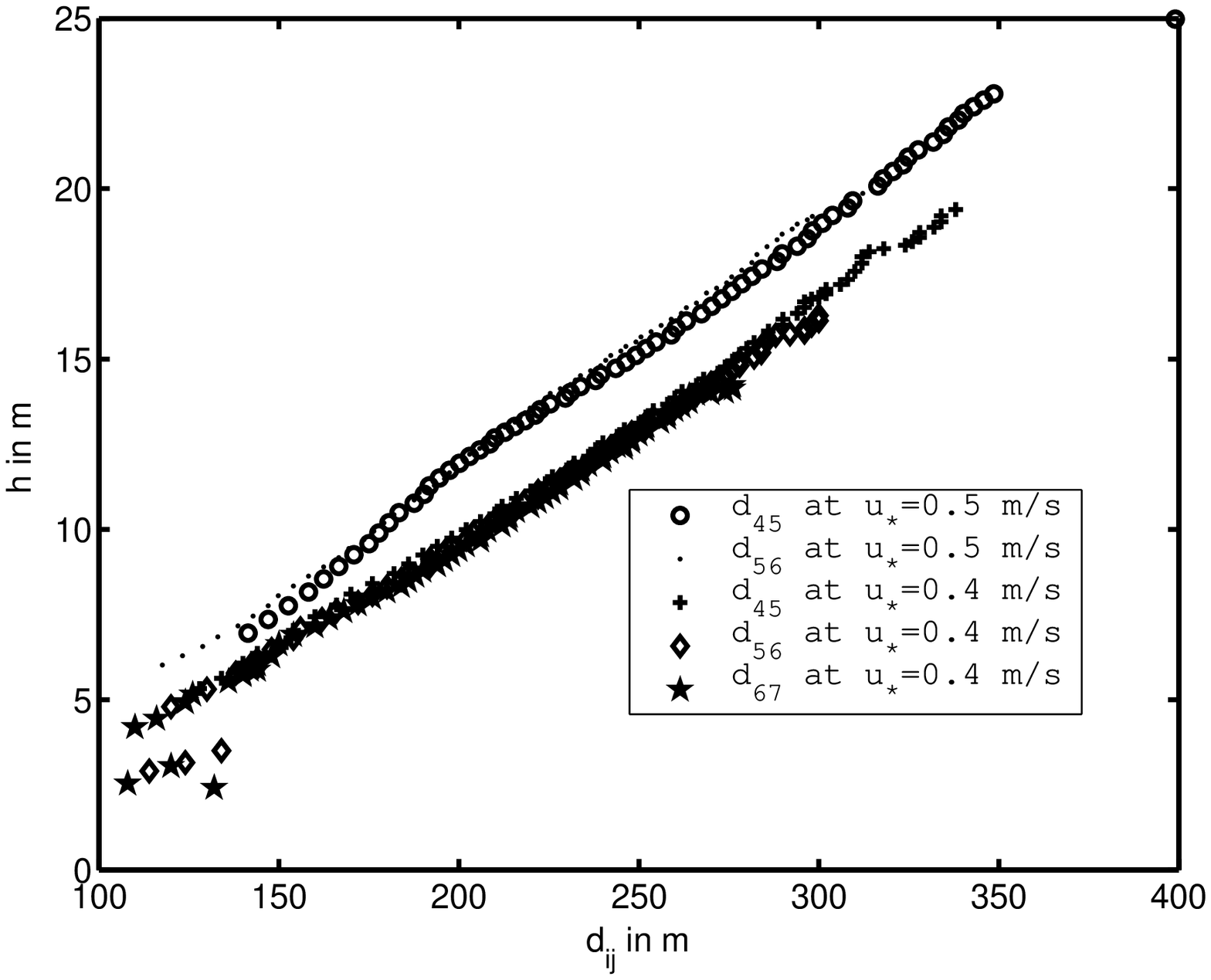}
    \caption{Relationship between spacing and height of the dunes for the shear velocities
             $u_* = 0.4$ ms$^{-1}$ and $u_* = 0.5$ ms$^{-1}$. The dependency seems to be linear.}
    \label{fig:transv_spach}
  \end{center}
\end{figure}

The results of the numerical calculations with different sand influxes $q_{in}$ at the same 
shear velocities lead to the conclusion that there is a direct dependency between influx
and height growth of the first dune (Figure~\ref{fig:transv_scalhq}). The nearer the sand 
influx gets to the saturation flux of saltation transport the slower increases the height
of the first dune. Hence, dune fields where the sand influx varies strongly in time around
the saturation flux of saltation initiate first dunes with different heights.
So there can be smaller dunes moving faster into bigger ones and 
the symmetry breaking explained in the Section~\ref{sec:trans_2dim} will occur.
In the following some relations found for the time evolution of transverse dunes are 
presented. 
Figure~\ref{fig:transv_scalhq} indicates also that height versus time increases with 
a power law as was also found in the model of \citeN{Momiji2001}, 
\begin{equation}
   h(t) \propto \sqrt{a \cdot t},
\end{equation}
where $a$ is a parameter which is dependent on shear velocity and sand influx. $a$ is a 
measure for the growth rate. The growth rate seems to be smaller for a larger distance from the
beginning of the dune field. There the dunes contain longer slip faces because of their
higher age. We observe from Figure~\ref{fig:transv_scalhq} that the growth rate should 
converge to a constant value for very large distances.
The same relation is found for the spacing $d_{ij}$ of the dunes $i$ and $j$ 
(Figure~\ref{fig:transv_scalsc}),
\begin{equation}
   d_{ij}(t) \propto \sqrt{b \cdot t},
\end{equation}
where $b$ denotes a parameter which measures the spacing rate. This spacing
rate approaches the same value for all dunes far away from the influx region. The values 
fitting well to these rates are
approximately $b=8.53 \cdot 10^{-5}$ m$^2$s$^{-1}$ and $b=1.56 \cdot 10^{-4}$ m$^2$s$^{-1}$ 
for a shear velocity $u_*=0.4$ m s$^{-1}$ and $u_*=0.5$ m s$^{-1}$, respectively.
Figure~\ref{fig:transv_spach} shows the spacing height relationship which seems to be linear.
This agrees with the data of measurements of transverse dunes in the Namib Sand Sea in Namibia
by \citeN{Lancaster83}. For different shear stresses the slope stays constant whereas the 
axis intercept of $h$ increases with higher shear velocities. Hence, the dunes in a transverse
dune field are located closer to each other for higher shear velocities. 

\subsection{Dune velocity}
\label{subsec:dunevel}

\begin{figure}[htb]
  \begin{center}
    \includegraphics[width=0.8\textwidth]{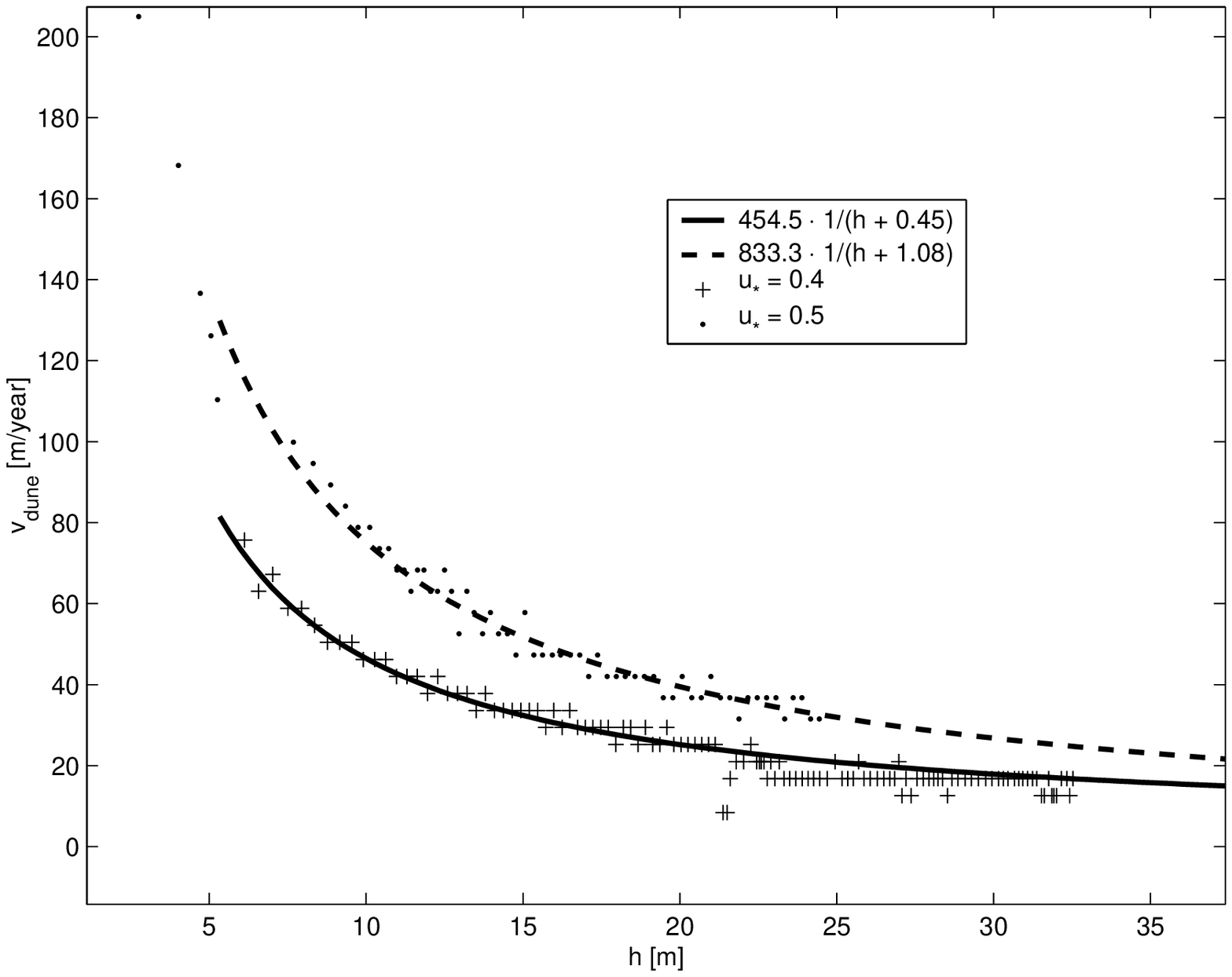}
    \caption{Dune velocity $v_{dune}$ versus height $h$ for 
             the shear stresses $u_*=0.4$ m s$^{-1}$ and $u_*=0.5$ m s$^{-1}$. The 
             velocity decreases proportional to the reciprocal height.}
    \label{fig:transv_scalv2}
  \end{center}
\end{figure}

The validity of Bagnold's law, 
\begin{equation}
   v_{dune} =  \frac{\Phi_{dune}}{h},
   \label{eq:transv_vdune}
\end{equation} 
where $\Phi_{dune}$ is the bulk flux of sand blown over the brink has been shown by 
observations of real dunes. According to 
\citeN{SauermannPhD2001} a better fit is given by using instead of the height $h$ the 
characteristic length $l$,
i.e. the length of the envelope comprising the height profile and the separation bubble.  
Bagnold's law already fits quite well. Nevertheless, the Equation~(\ref{eq:transv_vdune}) 
is generalized.
The length of the envelope can be expressed as a function of the height. The function is 
developed into a Taylor series and orders higher than the linear order are neglected. This 
finally yields, 
\begin{equation}
   v_{dune} =  \frac{\Phi_{dune}}{h+C},
   \label{eq:transv_vdune2}
\end{equation} 
where $C$ denotes a constant. 
In Figure~\ref{fig:transv_scalv2} the dune velocities with respect to the height and their
fits to Equation~(\ref{eq:transv_vdune2}) are compared
for the shear velocities $u_*=0.4$ m s$^{-1}$ and $u_*=0.5$ m s$^{-1}$. The observed bulk 
fluxes
are $\Phi_{dune}=454.5$ m$^2$s$^{-1}$ and $\Phi_{dune}=833.3$ m$^2$s$^{-1}$ with the 
corresponding constants $C=0.45$ m and $C=1.08$ m, respectively. The 
values are smaller than the bulk fluxes of isolated 2-dimensional dunes calculated by 
\citeN{SauermannPhD2001} on plain ground without sand. Thus the velocities are also smaller 
than those observed for isolated transverse dunes on plain ground without sand. 
\citeN{Lancaster85} found
less speed-up of the wind velocity over continuous sand dunes than over isolated transverse
dunes. This agrees with the results in this model. 

\section{The model of two dimensional dunes with periodic boundary}
\label{sec:trans_periodic}

The simulations in this section have a periodic boundary condition in wind direction. So the
parameter of sand influx that was used additionally before is no more available. Sand 
influx is set equal to the 
outflux. The avalanches flow down even if the slip face is divided by the boundary. 
Also the separation bubble follows the periodic boundary conditions. The calculations are made with dune 
fields of a length of two kilometers. The ground is completely filled up with sand.
The initial surface consists of small Gaussian hills which disturb a plain surface.

\subsection{Time evolution}

\begin{figure}[tb]
  \begin{center}
    \includegraphics[width=0.8\textwidth]{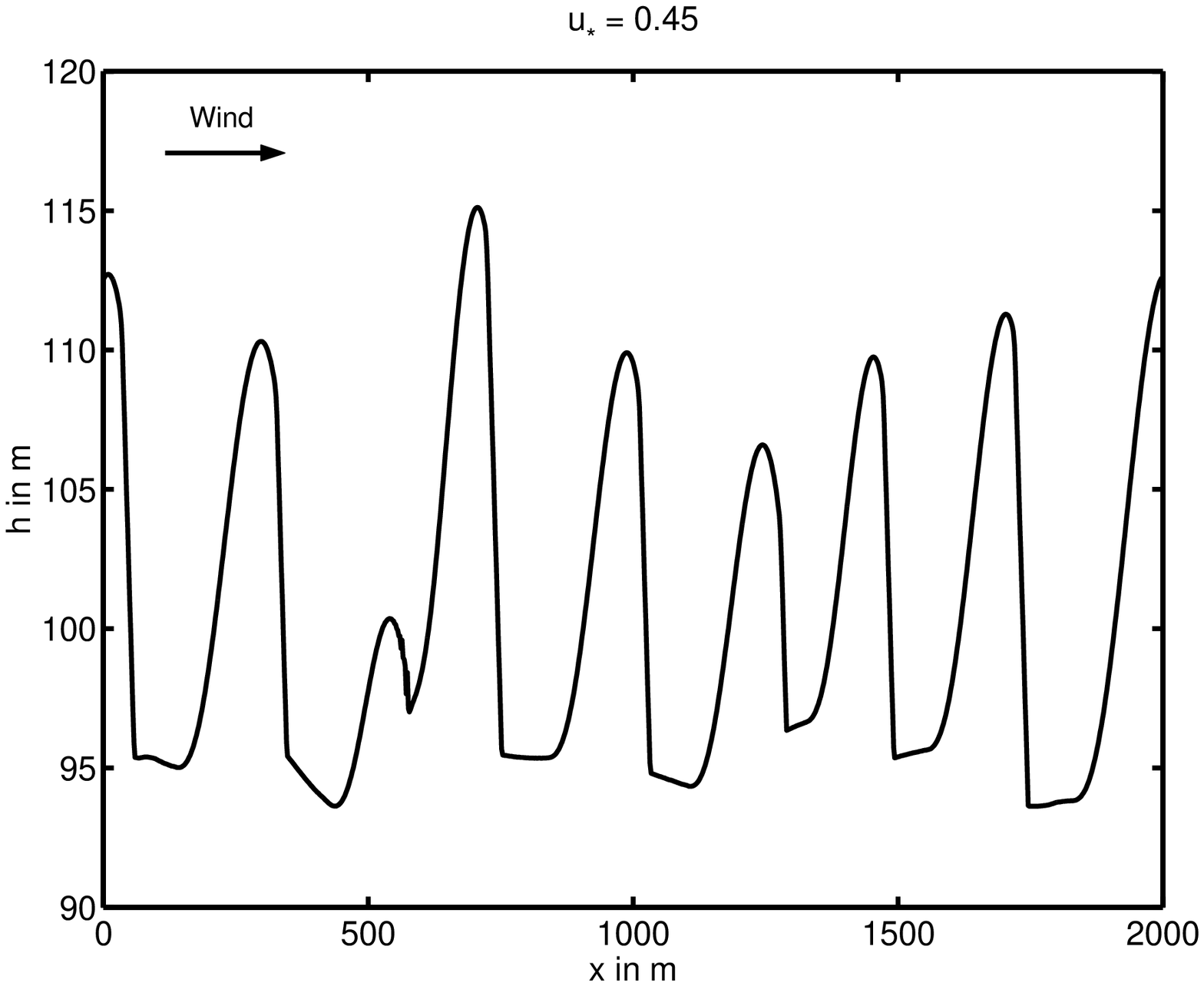}
    \caption{Surface of a dune field with a length of two kilometers. The shear velocity is 
             $u_*=0.45$ m s$^{-1}$ after $31.7$ years, the boundary conditions are periodic.
             The structure is 
             quite irregular.}
    \label{fig:transv_1dp1}
  \end{center}
\end{figure}

\begin{figure}[htb]
  \begin{center}
    \includegraphics[width=0.8\textwidth]{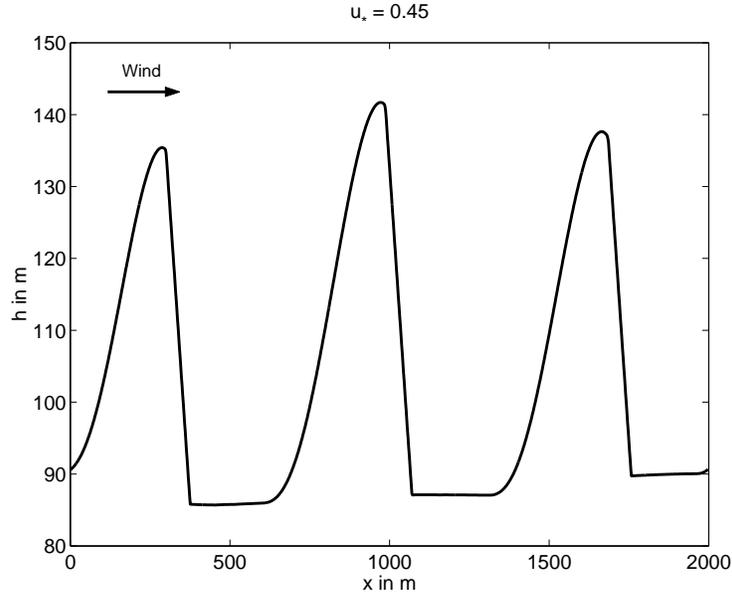}
    \caption{Surface of a dune field with a length of two kilometers. The shear velocity is 
             $u_*=0.45 $m s$^{-1}$ after $95.1$ years, the boundary conditions are periodic.
             There are three 
             dunes left with similar heights.}
    \label{fig:transv_1dp2}
  \end{center}
\end{figure}

\begin{figure}[htb]
  \begin{center}
    \includegraphics[width=0.8\textwidth]{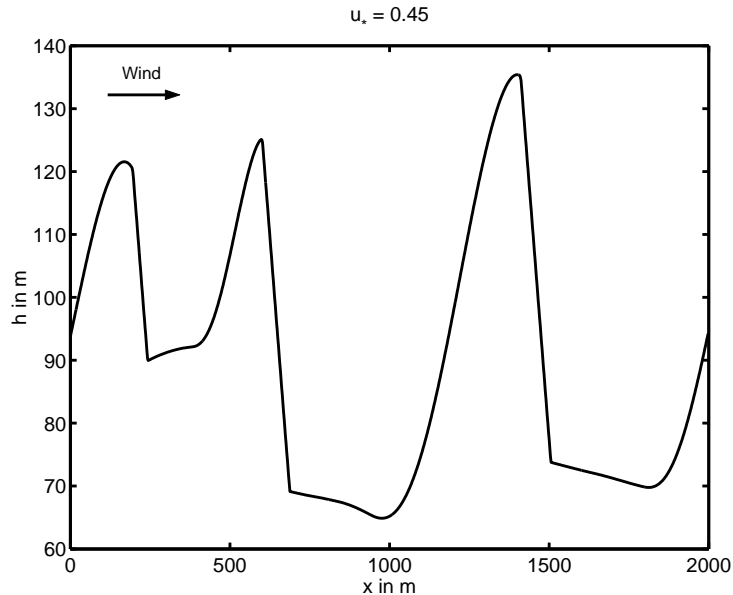}
    \caption{Surface of a dune field with a length of two kilometers. The shear velocity is 
             $u_*=0.45$ m s$^{-1}$ after $190.2$ years, the boundary conditions are periodic.
             The number of 
             dunes will decrease by one after coalescence of two dunes.}
    \label{fig:transv_1dp3}
  \end{center}
\end{figure}

\begin{figure}[tb]
  \begin{center}
    \includegraphics[width=0.8\textwidth]{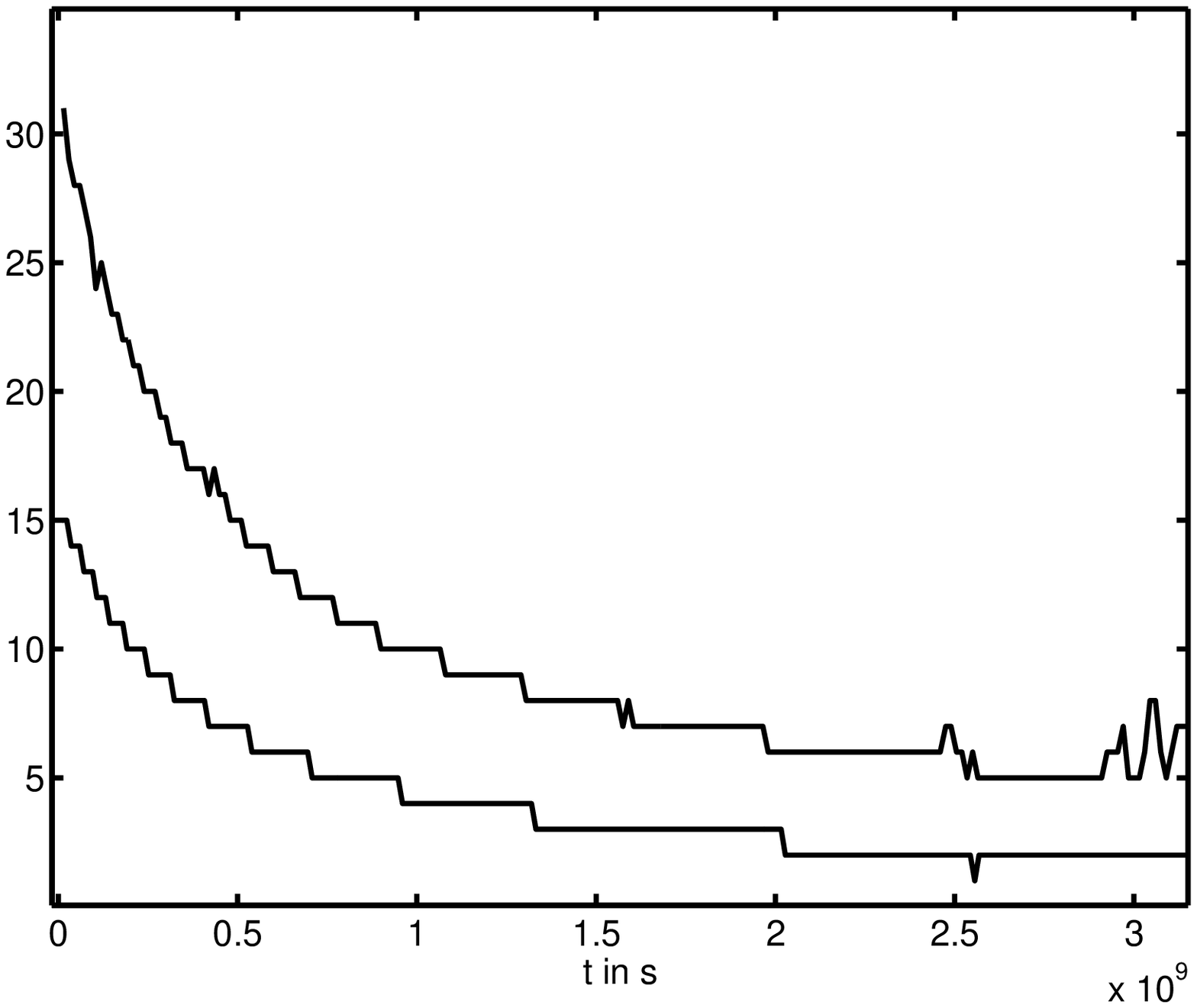}
    \caption{The number of dunes decreases quite regularly in time.}
    \label{fig:transv_nu_d}
  \end{center}
\end{figure}

The situation is similar to the simulations of  three dimensional transverse dune 
fields. In the beginning many dunes of different size grow in the system but after some time 
the dunes approach same heights. As in the  three dimensional case all the dunes keep 
growing so that the number of dunes has to decrease. A more detailed description of the 
process of merging dunes is given in the following section. The periodic boundary forces
a decrease of the number of dunes and this process makes the evolution rather
complex. 
Figures~\ref{fig:transv_1dp1}, \ref{fig:transv_1dp2} and \ref{fig:transv_1dp3} show the 
height profile of a 
dune field of a length of two kilometers at three time steps. Figure~\ref{fig:transv_1dp1}
depicts a surface with dunes of different sizes which seem to interact strongly with
each other. The regular state
of the system in Figure~\ref{fig:transv_1dp2} is disturbed in Figure~\ref{fig:transv_1dp3}.
The number of dunes decreases quite regularly versus time (Figure~\ref{fig:transv_nu_d}). This process is very slow
and the dunes grow much slower than in the model with open boundary.

\subsection{Do transversal dunes behave like solitons?}
\label{sec:trans_soliton}

\begin{figure}[htb]
  \begin{center}
    \includegraphics[width=0.8\textwidth]{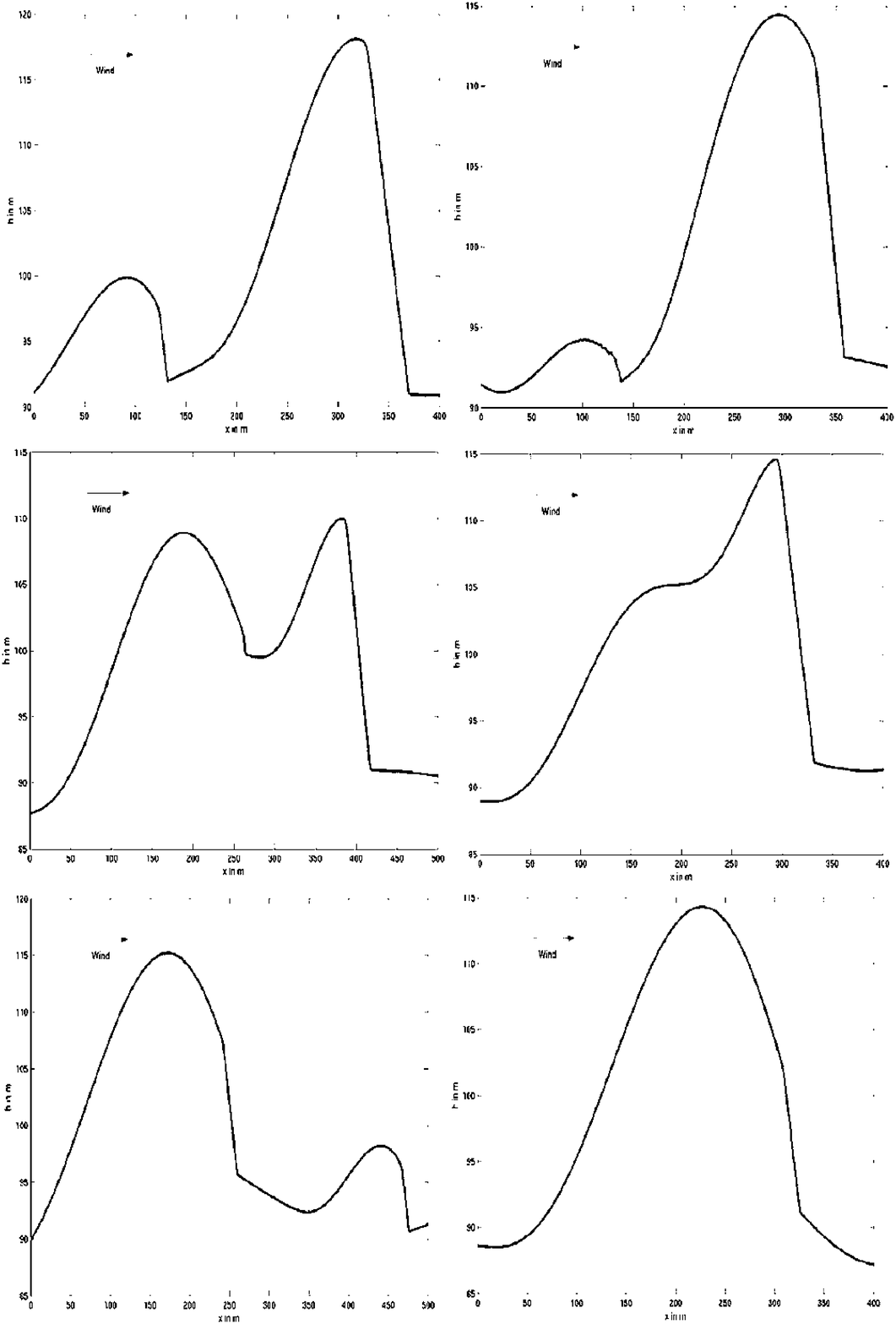}

    \caption{Left: Wandering of a small dune over a big one.
             Right: Coalescence of small dune in a big one.
             These results are part of a simulation of a transverse dune field with a length of 2km. 
             The shear velocity is $u_*=0.5 $m s$^{-1}$ and boundary conditions are 
             periodic.} 
    \label{fig:transv_sol1}
  \end{center}
\end{figure}

\citeN{Besler97} proposed that barchan dunes behave like solitons. Solitons are 
self-stabilizing wave packs which do not change their shape during their propagation
not even after collision with other waves. They are found in non-linear systems like 
for example in shallow water waves. Observations of barchan dunes seem to give evidence that small
dunes can migrate over bigger ones without being absorbed completely. 
A closer examination of the merging of two or more dunes from calculations of a 
two dimensional dune field with periodic boundary let to some interesting observations.
In this case the dune field cannot break its translational scale invariance to reach a faster
colliding of two adjacent dunes. Thus the supposition would be that a smaller dune due to 
its higher velocity collides with the next bigger one in wind direction without passing over
it. That is not the case. As an example see the Figure~\ref{fig:transv_sol1}. 
A small dune climbs up the windward side of
the following bigger dune. As it reaches the same height as the following one it seems to 
hand over the state of being the smaller dune. The dune that was bigger before then wanders 
towards the following dune of the dune field. This process can be 
observed several times. The fact that the volume of the smaller dune decreases after
every time it passes a dune let us conclude that there will be a final dune which will absorb the small 
one. Hence, these dunes do not behave exactly like solitons because of their loss of volume.
The suggestion of \citeN{Wilson72} that dunes of different hierarchies (for example 
dunes and mega dunes) could exist with each other without interactions is not
valid in our case. 
Nevertheless the fact that small dunes can migrate over others leads to the conclusion
that different hierarchies of dune sizes can coexist in a field of transverse dunes. 
The loss of volume demonstrates that there is some interaction between different
hierarchies of dune sizes. 

Finally Figure~\ref{fig:transv_sol1} shows a part of the same 
simulation where a very small dune finally swallowed the bigger one and the number of
dunes decreases by one.
%
%
%
%
%
\newpage

%
%
%
%
%
%
%

\section{Conclusions}
\label{sec:t_concl}

It was shown how transverse dune fields can develop with respect to time.
None of the simulations gave evidence that a final stage would be reached, 
neither the model with periodic nor the model with open boundaries. 
More knowledge about the still not very well understood trapping efficiency in the lee side of
the dunes and the inclusion of this effect as done by \citeN{MomijiWarren2000} could lead to stable 
final states. The use of translational invariance, 
shown in the three dimensional model, made it possible to restrict to the model of 
two dimensional dunes. All simulations also showed that the evolution of a slip face leads 
to a continuous growth of transversal dunes, and no final state will be reached. In the model 
of two dimensional dunes with constant
influx the height of the dunes increases proportional with respect to square root of time.
The same power law is found for the crest-to-crest spacing in the dune field.  
The difference of influx and saturation flux seems to play a crucial role in dune size.
The same relation between dune velocity and height 
is found for barchan dunes. Two dimensional modeling with periodic boundary shows
that different hierarchies of dunes with different sizes can exist but they interact with each
other.

\bibliographystyle{chicago}
\bibliography{dune}

\begin{thebibliography}{}

\bibitem[\protect\citeauthoryear{Bagnold}{Bagnold}{1941}]{Bagnold41}
Bagnold, R.~A. (1941).
\newblock {\em The physics of blown sand and desert dunes.}
\newblock London: Methuen.

\bibitem[\protect\citeauthoryear{Besler}{Besler}{1997}]{Besler97}
Besler, H. (1997).
\newblock Eine {W}anderd\"{u}ne als {S}oliton?
\newblock {\em {{P}hysikalische {B}l\"{a}etter.}\/}~{\em 10}, 983.

\bibitem[\protect\citeauthoryear{Bouchaud, Cates, Ravi~Prakash, and
  Edwards}{Bouchaud et~al.}{1994}]{Bouchaud94}
Bouchaud, J.~P., M.~E. Cates, J.~Ravi~Prakash, and S.~F. Edwards (1994).
\newblock Hysteresis and metastability in a continuum sandpile model.
\newblock {\em J. Phys. France I\/}~{\em 4}, 1383.

\bibitem[\protect\citeauthoryear{Breed and Grow}{Breed and
  Grow}{1979}]{Breed79}
Breed, C.~S. and T.~Grow (1979).
\newblock Morphology and distribution of dunes in sand seas observed by remote
  sensing.
\newblock {\em U.S. Geological Survey Professional Paper\/}~{\em 1052}, 266.

\bibitem[\protect\citeauthoryear{Burkinshaw and Rust}{Burkinshaw and
  Rust}{1993}]{Burkinshaw93}
Burkinshaw, J.R., I.~W. and I.~Rust (1993).
\newblock Wind--speed profiles over a reversing transverse dune.
\newblock {\em Geological Society Special Publication\/}~{\em 72}, 25--36.

\bibitem[\protect\citeauthoryear{Cooke, Warren, and Goudie}{Cooke
  et~al.}{1993}]{Cooke93}
Cooke, R., A.~Warren, and A.~Goudie (1993).
\newblock {\em Desert Geomorphology}.
\newblock London: UCL Press.

\bibitem[\protect\citeauthoryear{Fisher and Galdies}{Fisher and
  Galdies}{1988}]{Fisher88}
Fisher, P.~F. and P.~Galdies (1988).
\newblock A computer model for barchan-dune movement.
\newblock {\em Computer and Geosciences\/}~{\em 14-2}, 229--253.

\bibitem[\protect\citeauthoryear{Herrmann and Sauermann}{Herrmann and
  Sauermann}{2000}]{herrmann-sauermann:2000}
Herrmann, H.~J. and G.~Sauermann (2000).
\newblock The shape of dunes.
\newblock {\em Physica A\/}~{\em 283}, 24--30.

\bibitem[\protect\citeauthoryear{Kroy, Sauermann, and Herrmann}{Kroy
  et~al.}{2002}]{KroySauermann2002}
Kroy, K., G.~Sauermann, and H.~J. Herrmann (2002).
\newblock Minimal model for sand dunes.
\newblock {\em Phys. Rev. L.\/}~{\em 68}, 54301.

\bibitem[\protect\citeauthoryear{Lancaster}{Lancaster}{1982}]{Lancaster82}
Lancaster, N. (1982).
\newblock Dune on the skeleton coast, namibia: geomorphology and grain size
  relationships.
\newblock {\em Earth Surf. Processes\/}~{\em 7}, 575--587.

\bibitem[\protect\citeauthoryear{Lancaster}{Lancaster}{1983}]{Lancaster83}
Lancaster, N. (1983).
\newblock Controls of dune morphology in the namib sand sea.
\newblock {\em Eolian Sediments and Processes\/}, 261--289.

\bibitem[\protect\citeauthoryear{Lancaster}{Lancaster}{1985}]{Lancaster85}
Lancaster, N. (1985).
\newblock Variations in wind velocity and sand transport on the windward flanks
  of desert sand dunes.
\newblock {\em Sedimentology\/}~{\em 32}, 581--593.

\bibitem[\protect\citeauthoryear{Lettau and Lettau}{Lettau and
  Lettau}{1978}]{Lettau78}
Lettau, K. and H.~Lettau (1978).
\newblock Experimental and micrometeorological field studies of dune migration.
\newblock In H.~H. Lettau and K.~Lettau (Eds.), {\em Exploring the world's
  driest climate}, pp.\  110--147. Center for Climatic Research, Univ.
  Wisconsin: Madison.

\bibitem[\protect\citeauthoryear{McKenna~Neuman and Nickling}{McKenna~Neuman
  and Nickling}{2000}]{McKenna2000}
McKenna~Neuman, C., L.~N. and W.~Nickling (2000).
\newblock The effect of unsteady winds on sediment transport on the stoss slope
  of a transversal dune, silverpeak, nv, usa.
\newblock {\em Sedimentology\/}~{\em 112}, 211--26.

\bibitem[\protect\citeauthoryear{Momiji}{Momiji}{2001}]{Momiji2001}
Momiji, H. (2001).
\newblock {\em Mathematical modelling of the dynamics and morphology of aeolian
  dunes and dune fields}.
\newblock Ph.\ D. thesis, University of London.

\bibitem[\protect\citeauthoryear{Momiji and Warren}{Momiji and
  Warren}{2000}]{MomijiWarren2000}
Momiji, H. and A.~Warren (2000).
\newblock Relations of sand trapping efficiency and migration speed of
  transverse dunes to wind velocity.
\newblock {\em Earth Surface Processes and Landforms\/}~{\em 25}, 1069--1084.

\bibitem[\protect\citeauthoryear{Mulligan}{Mulligan}{1988}]{Mulligan88}
Mulligan, K. (1988).
\newblock Velocity profiles measured on the windward slope of a transverse
  dune.
\newblock {\em Earth Surface Processes and Landforms\/}~{\em 13}, 573--582.

\bibitem[\protect\citeauthoryear{Nishimori, Yamasaki, and Andersen}{Nishimori
  et~al.}{1999}]{NishimoriXX}
Nishimori, H., M.~Yamasaki, and K.~H. Andersen (1999).
\newblock A simple model for the various pattern dynamics of dunes.
\newblock {\em Int. J. of Modern Physics B\/}~{\em 12}, 257--272.

\bibitem[\protect\citeauthoryear{Sauermann}{Sauermann}{2001}]{SauermannPhD2001}
Sauermann, G. (2001).
\newblock {\em Modeling of Wind Blown Sand and Desert Dunes}.
\newblock Ph.\ D. thesis, University of Stuttgart.

\bibitem[\protect\citeauthoryear{Sauermann, Kroy, and Herrmann}{Sauermann
  et~al.}{2001}]{SauermannKroy2001}
Sauermann, G., K.~Kroy, and H.~J. Herrmann (2001).
\newblock A continuum saltation model for sand dunes.
\newblock {\em Phys. Rev. E\/}~{\em 64}, 31305.

\bibitem[\protect\citeauthoryear{S{\o}rensen}{S{\o}rensen}{1991}]{Sorensen91}
S{\o}rensen, M. (1991).
\newblock An analytic model of wind-blown sand transport.
\newblock {\em Acta Mechanica (Suppl.)\/}~{\em 1}, 67--81.

\bibitem[\protect\citeauthoryear{Stam}{Stam}{1997}]{Stam97}
Stam, J. M.~T. (1997).
\newblock On the modelling of two-dimensional aeolian dunes.
\newblock {\em Sedimentology\/}~{\em 44}, 127--141.

\bibitem[\protect\citeauthoryear{van Boxel, Arens, and van Dijk}{van Boxel
  et~al.}{1999}]{boxel-arens-van_dijk:99}
van Boxel, J.~H., S.~M. Arens, and P.~M. van Dijk (1999).
\newblock Aeolian processes across transverse dunes i: Modelling the air flow.
\newblock {\em Earth Surf. Process. Landforms\/}~{\em 24}, 255--270.

\bibitem[\protect\citeauthoryear{van Dijk, Arens, and van Boxel}{van Dijk
  et~al.}{1999}]{van_dijk-arens-boxel:99}
van Dijk, P.~M., S.~M. Arens, and J.~H. van Boxel (1999).
\newblock Aeolian processes across transverse dunes ii: Modelling the sediment
  transport and profile development.
\newblock {\em Earth Surf. Process. Landforms\/}~{\em 24}, 319--333.

\bibitem[\protect\citeauthoryear{Weng, Hunt, Carruthers, Warren, Wiggs,
  Livingstone, and Castro}{Weng et~al.}{1991}]{weng-etal:91}
Weng, W.~S., J.~C.~R. Hunt, D.~J. Carruthers, A.~Warren, G.~F.~S. Wiggs,
  I.~Livingstone, and I.~Castro (1991).
\newblock Air flow and sand transport over sand--dunes.
\newblock {\em Acta Mechanica (Suppl.)\/}~{\em 2}, 1--22.

\bibitem[\protect\citeauthoryear{Wilson}{Wilson}{1972}]{Wilson72}
Wilson, I. (1972).
\newblock Aeolian bedforms--their development an origins.
\newblock {\em Sedimentology\/}~{\em 19}, 173--210.

\bibitem[\protect\citeauthoryear{Wippermann and Gross}{Wippermann and
  Gross}{1986}]{Wippermann86}
Wippermann, F.~K. and G.~Gross (1986).
\newblock The wind-induced shaping and migration of an isolated dune: A
  numerical experiment.
\newblock {\em Boundary Layer Meteorology\/}~{\em 36}, 319--334.

\bibitem[\protect\citeauthoryear{Zeman and Jensen}{Zeman and
  Jensen}{1988}]{zeman-jensen:88}
Zeman, O. and N.~O. Jensen (1988).
\newblock Progress report on modeling permanent form sand dunes.
\newblock {\em Ris{\o} National Laboratory\/}~{\em M-2738}.

\end{thebibliography}
\end{document}